\documentstyle[11pt,conf_iap,epsf]{article}
\begin{document}
\heading{CEPHEIDS IN THE LMC: RESULTS FROM THE MACHO PROJECT}

\photo{ }

\author{D.L.~WELCH$^{1}$, 
C.~ALCOCK$^{2}$, 
R.A.~ALLSMAN$^{3}$, 
D.R.~ALVES$^{2}$,
T.S.~AXELROD$^{3}$, 
A.~BECKER$^{5}$, 
D.P.~BENNETT$^{2}$, 
K.H.~COOK$^{2}$, 
K.C.~FREEMAN$^{3}$,
K.~GRIEST$^{4}$, 
J.A.~GUERN$^{4}$,
M.J.~LEHNER$^{4}$, 
S.L.~MARSHALL$^{2}$, 
D.~MINNITI$^{2}$,
B.A.~PETERSON$^{3}$, 
M.R.~PRATT$^{5}$, 
P.J.~QUINN$^{6}$, 
A.W.~RODGERS$^{3}$, 
A.~RORABECK$^{1}$,
C.W.~STUBBS$^{5}$,
W.~SUTHERLAND$^{7}$}
{$^{1}$ Dept. of Physics and Astronomy, McMaster Univ., Hamilton, ON, L8S 4M1, 
Canada,
$^{2}$\it Lawrence Livermore National Laboratory, Livermore, CA94550,
U.S.A., \\
$^{3}$\it Mt Stromlo and Siding Spring Observatories, Australian National
Univ., ACT2611, Australia, \\
$^{4}$ Dept. of Physics, Univ. of California, San Diego, CA92093, U.S.A.,\\
$^{5}$ Dept. of Astronomy, Univ. of Washington, Seattle, WA98195, U.S.A.,\\
$^{6}$ European Southern Observatory, Karl-Schwarzchild Str. 2, D-85748, 
Garching, Germany, \\
$^{7}$ Dept. of Physics, Univ. of Oxford, Oxford, OX1 3RH, U.K.}

\bigskip
\begin{abstract}{\baselineskip 0.4cm
The approximately 1500 Cepheid variables in 22 LMC fields of the MACHO 
Project survey have been calibrated and analysed. In this paper, we report 
improved period ratios for a total of 73 beat Cepheids and provide a first
look at the Fourier decomposition parameters for both singly- and doubly-periodic
Cepheids. We also note some an unusual amplitude-changing pulsator.
}
\end{abstract}

\section{Introduction}

The MACHO Project has been collecting photometry of stars in LMC fields for 
over 1400 days at this writing and has already provided a wealth of important
new information on variable stars - see \cite{AlcockBC},\cite{AlcockRRd1},\cite{AlcockRCrB},
\cite{AlcockRRd2}. There is, naturally, great interest in the classical Cepheid
variables which are used to calibrate the extragalactic distance scale. At present
our Cepheid catalogue contains 1466 single-mode (classical) and 73 double-mode (beat)
Cepheids. Most of these stars have between 300 and 1100 individual epochs of two-color
photometry. We have determined periods for all stars using the full dataset and
have generated finder charts, Fourier amplitudes, phases, and ratios and will submit
a full analysis of these results in the near future. This paper contains a brief
summary of interesting intermediate results, including a list of improved period
ratios for LMC beat Cepheids.

\section{Fourier Decomposition}

The properties of lightcurve shapes are observables which are not
affected by reddening or distance. Furthermore, both the photometry
and period of light variation can usually be determined with high
precision. Therefore, it is of considerable interest to investigate
whether lightcurve shape is correlated with other physical quantities
that are, in practice, more difficult to determine, such as position
in the instability strip, pulsation mode, and metallicity.

Fourier coefficients for LMC Cepheids have been most recently discussed
by \cite{Beaulieu1995}. As a by-product of a search for evidence of 
microlensing toward the LMC, the EROS collaboration
detected 72 Cepheids in their 0.4 square degree field. Approximately 1000
epochs of photometry in the B and R bandpasses were obtained. They discuss
the low-order Fourier coefficients so derived.

\begin{figure}
\begin{center}
\begin{minipage}{12cm}
\epsfxsize=12cm
\epsfbox{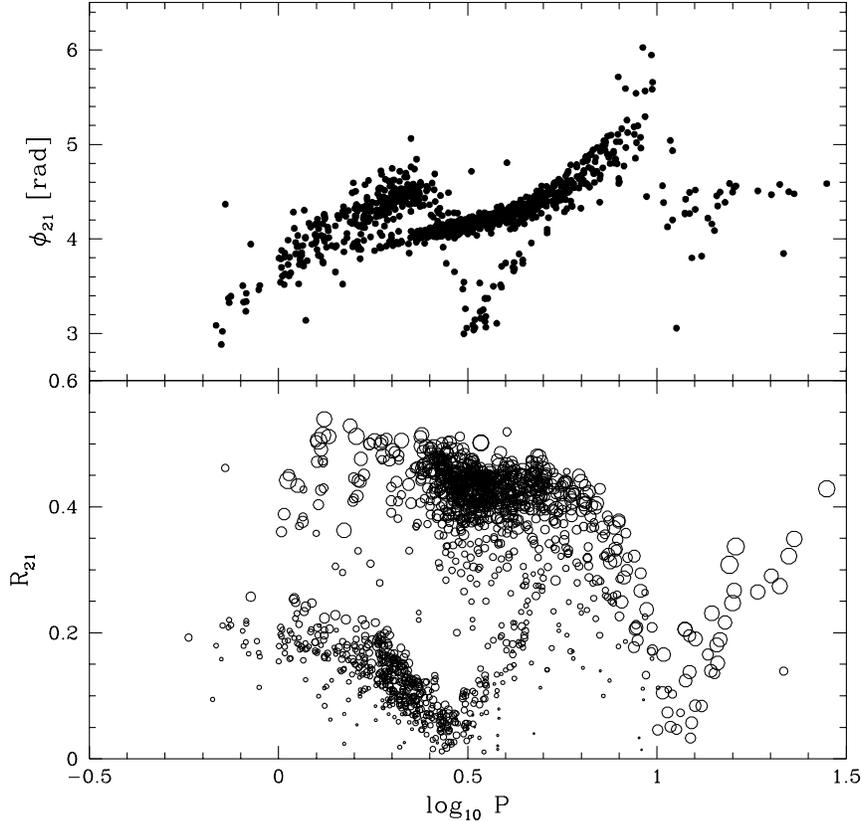}
\end{minipage}
\caption{The Fourier amplitude ratios and phase differences for stars with
phase difference uncertainty less than 0.1 rad and amplitude ratio uncertainty
less than 0.05 in the $V$ bandpass. The size of the amplitude ratio symbol
is proportional to the amplitude of the lowest order amplitude.}
\label{figfour21}
\end{center}
\end{figure}

We have undertaken Fourier decomposition of our lightcurves for order n=12
using the technique described by \cite{SL1981}. The quality of our Fourier
fits has been significantly improved by the length of the time series - now
three to four times as long as that reported in \cite{Welch1995}. The division
between overtone and fundamental Cepheids is now even more distinct and there
is also an obvious dependance of $R_{21}$ on the amplitude $R_1$. The so-called
long-period $s$ Cepheid sequence is also better defined. Examples of the run
of Fourier parameters with photometric period are given in Figure~\ref{figfour21}.
The 2:1 resonance is clearly located by this data at $\log_{10}$ P = 1.05 $\pm$ 0.03.

The run of Fourier parameters with period defined by this sample will ultimately
be useful in defining a simple parameterization of lightcurve shape with period,
peak-to-peak amplitude and bandpass.

\section{Beat Cepheids}

\begin{figure}
\begin{center}
\begin{minipage}{10cm}
\epsfxsize=10cm
\leavevmode
\epsfbox{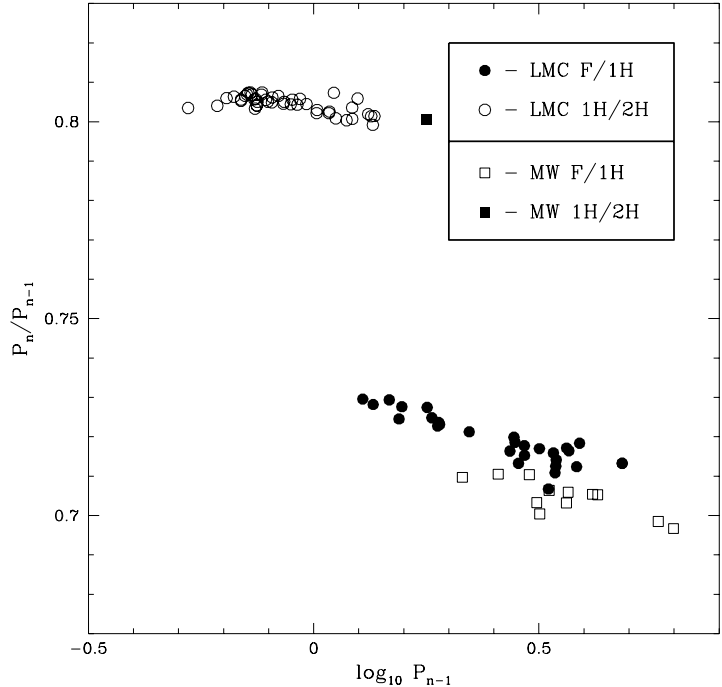}
\end{minipage}
\caption{The Petersen diagram for 73 beat Cepheids in the LMC and 14 in the Galaxy.}
\label{figbeatrat}
\end{center}
\end{figure}

\begin{figure}
\begin{center}
\begin{minipage}{10cm}
\epsfxsize=10cm
\epsfbox{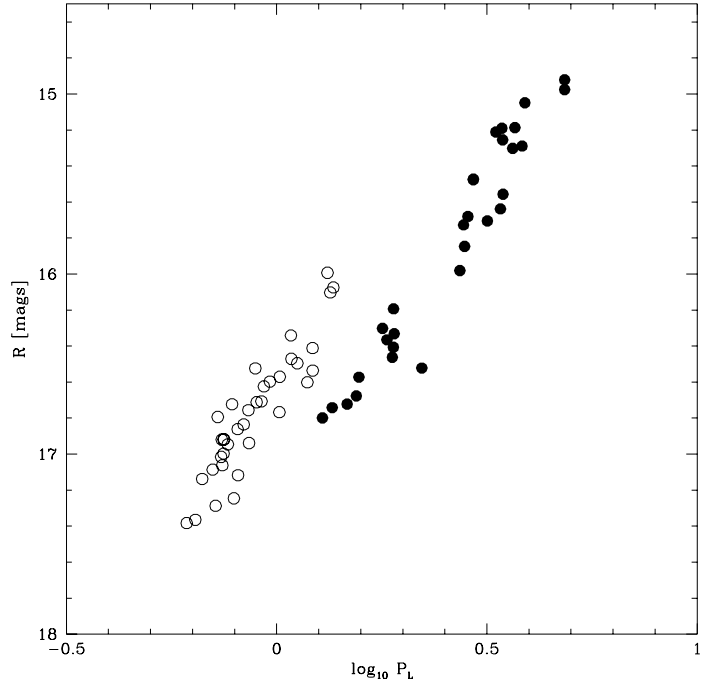}
\end{minipage}
\caption{The Period-Luminosity relation for 73 beat Cepheids in the LMC. The symbols
have the same meaning as in Figure~\ref{figbeatrat}.}
\label{figpl}
\end{center}
\end{figure}

The metallicity sensitivity of the period ratios of beat Cepheids make them
especially interesting. The sample of 45 stars reported in \cite{AlcockBC} has
been enlarged and refined by the addition of two more years of observations.
The periods and period ratios reported here were obtained by a simultaneous 
weighted least-squares fit to the principal frequencies and their harmonics, and
mixing terms up to order 3.
Table 1 contains improved period ratios for beat Cepheids in the LMC. The columns,
from left to right, are the designation using equinox J2000.0 coordinates, the
internal identifier, the longest principal period in days, the ratio of the two
periods present, and the mode identification. Additional data on these stars will
be presented in \cite{AlcockBC2}. We note that there are 75 entries in this table
but that two of the stars are found in overlap regions of different fields and
so the total number of distinct stars is 73. A plot of period ratio versus period
for the 75 entries is shown in Figure~\ref{figbeatrat}. We also present the $R$-band
P-L relation using the longer of the two principal periods in Figure~\ref{figpl}.

\begin{figure}
\begin{center}
\begin{minipage}{10cm}
\epsfxsize=10cm
\leavevmode
\epsfbox{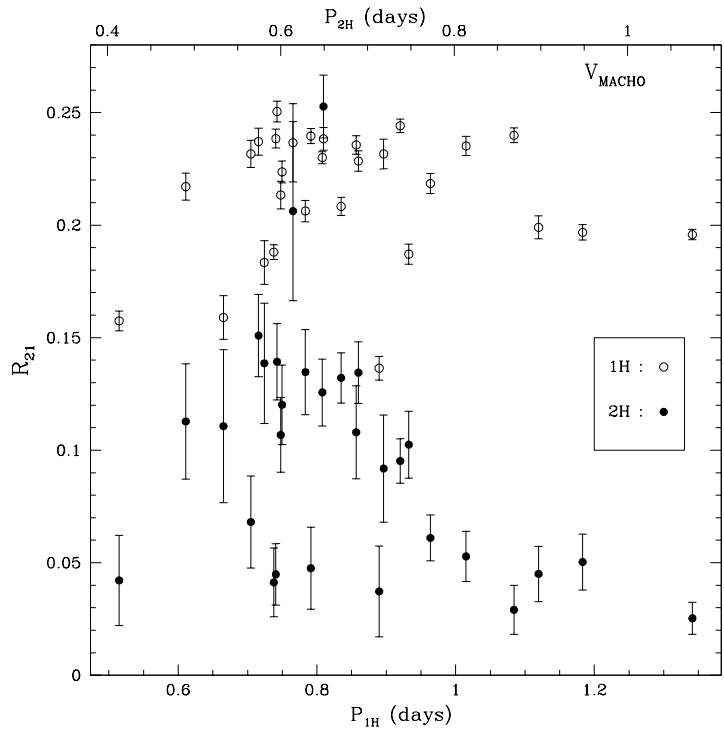}
\end{minipage}
\caption{$R_{21}$ for the instrumental MACHO $V$ band.  Note the two 
distinct sequences, as well as the decrease of the 2H mode as the 2H period 
approaches 1.2 days.}
\label{rtovsp}
\end{center}
\end{figure}

\begin{figure}
\begin{center}
\begin{minipage}{10cm}
\epsfxsize=10cm
\leavevmode
\epsfbox{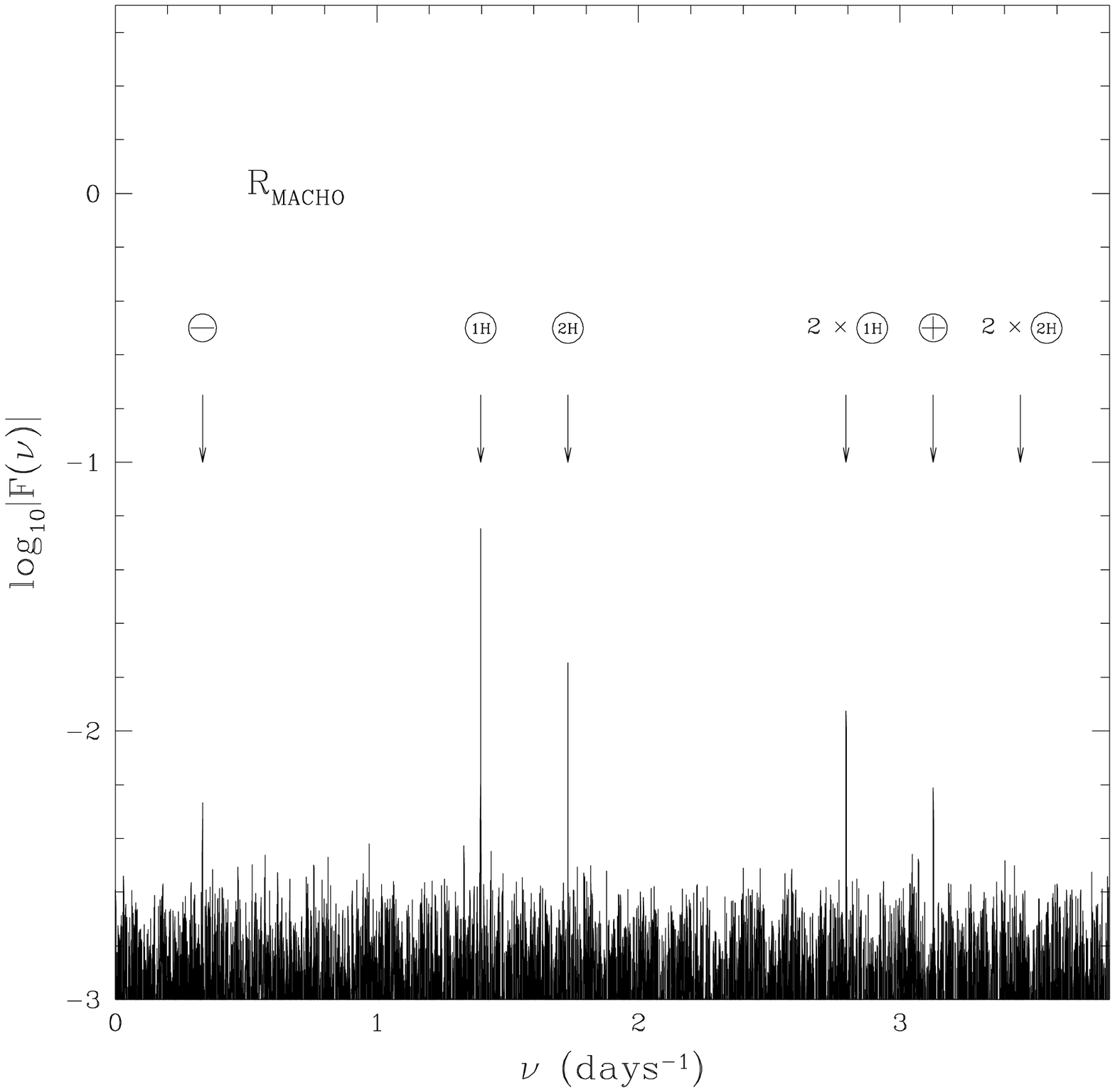}
\end{minipage}
\caption{A typical 1H/2H beat Cepheid power spectrum (80.7080.2618) in the 
instrumental MACHO $R$ band. The 1H and 2H frequencies and their harmonics are 
labelled, as are the sum and difference frequencies.  Note the abscence of any 
2H harmonics, which is typical for these stars.}
\label{powspec}
\end{center}
\end{figure}

The lightcurve geometry of the 2H mode can be extracted from the existing 1H/2H
lightcurves. We find that there is no significant power present at any of the
harmonics of the 2H period. The Fourier amplitude ratio $R_{21}$ is shown in
Figure~\ref{rtovsp} as a function of period. Note the two distinct sequences.  
$R_{21}$ seems to approach zero for the 2H mode when $P_{2H}$ is about 1.2 
days, which allows us to confine our search for singly-periodic 2H to below this
period. The power spectrum for a typical 1H/2H Cepheid is shown in 
Figure~\ref{powspec}.

\section{Unusual Cepheids}

\begin{figure}
\begin{center}
\begin{minipage}{10cm}
\epsfxsize=10cm
\epsfbox{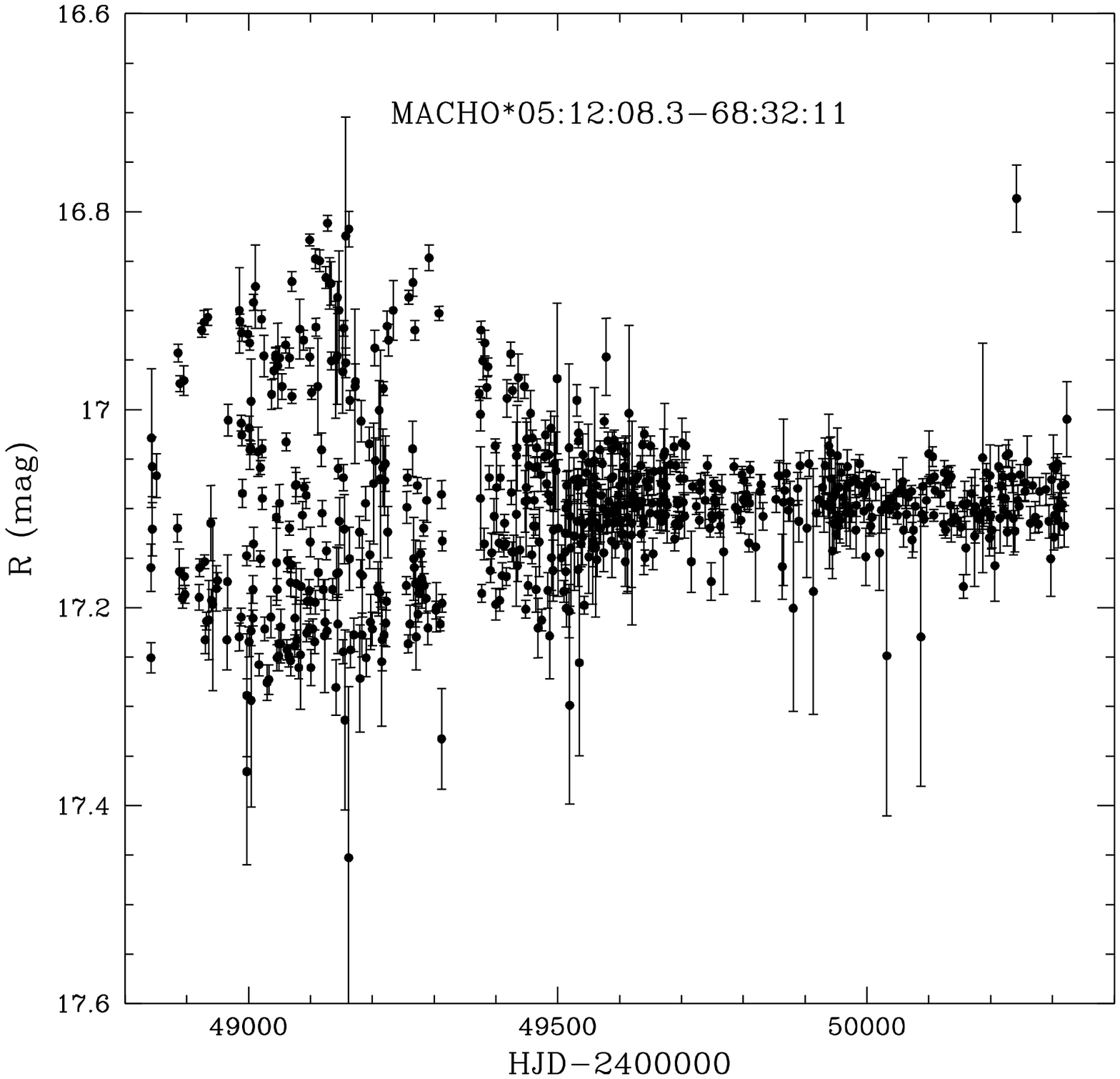}
\end{minipage}
\caption{The lightcurve of {\tt MACHO*05:12:08.3-68:32:11} during an interval
of 1480 days. This star has a period of 0.763 days and may be related to the RR Lyrae
stars. 
}
\label{figampchange}
\end{center}
\end{figure}

We have found a number of examples of stars which are apparently changing their
amplitudes. The most spectacular of these is {\tt MACHO*05:12:08.3-68:32:11} which
has a period of 0.763 days and may be an RR Lyrae star. The $R$ lightcurve for this star 
is shown in Figure~\ref{figampchange}.

\newpage

\begin{center}
{\bf Table 1.} Improved LMC Beat Cepheid Period Ratios
\end{center}
\begin{center}
\begin{tabular}{|c|r|c|c|c|}
\hline
&&&&\\
Cepheid&Identifier\ \ \ \ \ &$P_{n-1}$&$P_{n}$/$P_{n-1}$&Modes\\
&&{\sl (days)}&&\\
\hline 
&&&&\\
{\tt MACHO*05:11:39.9$-$68:49:58}&{\tt 79..5022...339}&{\tt 0.526263(000)}&{\tt 0.803491(007)}&{\tt 1H/2H}\\
{\tt MACHO*05:38:51.1$-$69:49:23}&{\tt 81..9485....45}&{\tt 0.610977(003)}&{\tt 0.804052(014)}&{\tt 1H/2H}\\
{\tt MACHO*05:48:55.2$-$70:30:02}&{\tt 12.11168...150}&{\tt 0.640693(004)}&{\tt 0.805979(018)}&{\tt 1H/2H}\\
{\tt MACHO*05:15:35.2$-$68:57:07}&{\tt 79..5747...197}&{\tt 0.665186(002)}&{\tt 0.806371(008)}&{\tt 1H/2H}\\
{\tt MACHO*05:15:05.4$-$69:39:55}&{\tt 5..5615..2564}&{\tt 0.689981(002)}&{\tt 0.805317(012)}&{\tt 1H/2H}\\
{\tt MACHO*05:15:06.3$-$69:39:54}&{\tt 78..5615....82}&{\tt 0.689979(003)}&{\tt 0.805597(013)}&{\tt 1H/2H}\\
{\tt MACHO*05:35:11.1$-$70:17:05}&{\tt 11..8873...189}&{\tt 0.704754(002)}&{\tt 0.806546(007)}&{\tt 1H/2H}\\
{\tt MACHO*05:16:28.6$-$69:36:33}&{\tt 78..5858...301}&{\tt 0.711553(004)}&{\tt 0.806921(014)}&{\tt 1H/2H}\\
{\tt MACHO*05:24:13.9$-$68:49:34}&{\tt 80..7080..2618}&{\tt 0.715870(002)}&{\tt 0.807350(006)}&{\tt 1H/2H}\\
{\tt MACHO*05:30:01.9$-$69:11:39}&{\tt 82..8042...142}&{\tt 0.724083(003)}&{\tt 0.807404(008)}&{\tt 1H/2H}\\
&&&&\\
{\tt MACHO*05:26:02.1$-$69:52:09}&{\tt 77..7427...306}&{\tt 0.729244(004)}&{\tt 0.806918(012)}&{\tt 1H/2H}\\
{\tt MACHO*05:37:47.1$-$70:51:43}&{\tt 11..9348....78}&{\tt 0.737786(001)}&{\tt 0.805670(008)}&{\tt 1H/2H}\\
{\tt MACHO*05:26:01.5$-$69:30:42}&{\tt 77..7432...248}&{\tt 0.740810(002)}&{\tt 0.803312(006)}&{\tt 1H/2H}\\
{\tt MACHO*04:58:53.0$-$68:51:07}&{\tt 18..2965...104}&{\tt 0.742535(004)}&{\tt 0.805778(012)}&{\tt 1H/2H}\\
{\tt MACHO*05:16:28.5$-$69:25:35}&{\tt 79..5861..5053}&{\tt 0.747883(002)}&{\tt 0.804089(005)}&{\tt 1H/2H}\\
{\tt MACHO*05:16:28.5$-$69:25:35}&{\tt 78..5861...239}&{\tt 0.747873(004)}&{\tt 0.804132(010)}&{\tt 1H/2H}\\
{\tt MACHO*05:34:31.9$-$69:45:15}&{\tt 81..8760...204}&{\tt 0.749790(004)}&{\tt 0.805021(013)}&{\tt 1H/2H}\\
{\tt MACHO*05:15:49.4$-$68:41:52}&{\tt 2..5750..2010}&{\tt 0.765462(005)}&{\tt 0.806873(011)}&{\tt 1H/2H}\\
{\tt MACHO*05:09:29.8$-$68:21:20}&{\tt 2..4788....82}&{\tt 0.768301(002)}&{\tt 0.807441(011)}&{\tt 1H/2H}\\
{\tt MACHO*05:23:13.8$-$69:36:36}&{\tt 78..6947..2839}&{\tt 0.783376(003)}&{\tt 0.805540(013)}&{\tt 1H/2H}\\
&&&&\\
{\tt MACHO*05:34:28.5$-$68:57:01}&{\tt 82..8772....88}&{\tt 0.791229(002)}&{\tt 0.805080(008)}&{\tt 1H/2H}\\
{\tt MACHO*05:38:45.4$-$70:36:11}&{\tt 11..9473...117}&{\tt 0.807845(003)}&{\tt 0.804919(016)}&{\tt 1H/2H}\\
{\tt MACHO*05:21:25.4$-$69:52:52}&{\tt 78..6701...236}&{\tt 0.809516(008)}&{\tt 0.806159(015)}&{\tt 1H/2H}\\
{\tt MACHO*05:23:59.1$-$69:15:30}&{\tt 80..7073...142}&{\tt 0.834857(002)}&{\tt 0.806534(006)}&{\tt 1H/2H}\\
{\tt MACHO*05:34:34.6$-$70:18:20}&{\tt 11..8751...129}&{\tt 0.856631(005)}&{\tt 0.804536(016)}&{\tt 1H/2H}\\
{\tt MACHO*05:46:42.8$-$70:40:50}&{\tt 12.10803...112}&{\tt 0.859789(006)}&{\tt 0.805044(013)}&{\tt 1H/2H}\\
{\tt MACHO*05:24:33.2$-$70:09:30}&{\tt 7..7181..1511}&{\tt 0.889713(007)}&{\tt 0.804411(016)}&{\tt 1H/2H}\\
{\tt MACHO*05:21:16.6$-$69:52:00}&{\tt 78..6580...150}&{\tt 0.896286(006)}&{\tt 0.805608(016)}&{\tt 1H/2H}\\
{\tt MACHO*05:23:10.0$-$70:28:45}&{\tt 6..6934....67}&{\tt 0.920147(002)}&{\tt 0.804300(006)}&{\tt 1H/2H}\\
{\tt MACHO*05:30:11.7$-$69:52:02}&{\tt 77..8032...175}&{\tt 0.932506(003)}&{\tt 0.805776(008)}&{\tt 1H/2H}\\
&&&&\\
{\tt MACHO*05:49:27.6$-$71:32:07}&{\tt 15.11153....34}&{\tt 0.963677(004)}&{\tt 0.804477(009)}&{\tt 1H/2H}\\
{\tt MACHO*05:47:11.7$-$70:41:10}&{\tt 12.10803....77}&{\tt 1.015097(010)}&{\tt 0.802180(016)}&{\tt 1H/2H}\\
{\tt MACHO*05:07:44.6$-$68:35:20}&{\tt 19..4421...403}&{\tt 1.017522(012)}&{\tt 0.802970(021)}&{\tt 1H/2H}\\
{\tt MACHO*05:21:05.4$-$68:23:35}&{\tt 3..6602....41}&{\tt 1.081242(006)}&{\tt 0.802169(012)}&{\tt 1H/2H}\\
{\tt MACHO*05:10:15.3$-$68:20:28}&{\tt 2..4909....67}&{\tt 1.084074(006)}&{\tt 0.802568(012)}&{\tt 1H/2H}\\
{\tt MACHO*05:25:59.2$-$69:49:13}&{\tt 77..7428...149}&{\tt 1.108829(035)}&{\tt 0.807335(033)}&{\tt 1H/2H}\\
{\tt MACHO*05:45:22.1$-$70:50:12}&{\tt 12.10558...923}&{\tt 1.119640(007)}&{\tt 0.800828(011)}&{\tt 1H/2H}\\
{\tt MACHO*05:43:20.9$-$71:08:48}&{\tt 15.10191....50}&{\tt 1.183152(010)}&{\tt 0.800353(023)}&{\tt 1H/2H}\\
{\tt MACHO*05:09:08.0$-$68:56:43}&{\tt 1..4658....66}&{\tt 1.217551(006)}&{\tt 0.803593(014)}&{\tt 1H/2H}\\
{\tt MACHO*05:07:37.0$-$69:12:47}&{\tt 1..4412...130}&{\tt 1.217977(005)}&{\tt 0.800710(009)}&{\tt 1H/2H}\\
&&&&\\
\hline
\end{tabular}
\end{center}
\begin{center}
\begin{tabular}{|c|r|c|c|c|}
\hline
&&&&\\
Cepheid&Identifier\ \ \ \ \ &$P_{n-1}$&$P_{n}$/$P_{n-1}$&Modes\\
&&{\sl (days)}&&\\
\hline
&&&&\\
{\tt MACHO*05:49:28.9$-$70:22:40}&{\tt 12.11170....25}&{\tt 1.251631(005)}&{\tt 0.805851(020)}&{\tt 1H/2H}\\
{\tt MACHO*05:39:29.3$-$70:38:20}&{\tt 11..9593....90}&{\tt 1.285135(010)}&{\tt 0.729568(007)}&{\tt F/1H}\\
{\tt MACHO*05:02:09.7$-$68:51:32}&{\tt 1..3570....55}&{\tt 1.321197(010)}&{\tt 0.801925(015)}&{\tt 1H/2H}\\
{\tt MACHO*05:20:19.7$-$70:42:29}&{\tt 13..6446....38}&{\tt 1.341420(008)}&{\tt 0.801375(014)}&{\tt 1H/2H}\\
{\tt MACHO*05:37:36.2$-$69:44:20}&{\tt 81..9244....71}&{\tt 1.352937(012)}&{\tt 0.799210(024)}&{\tt 1H/2H}\\
{\tt MACHO*05:16:23.4$-$68:35:26}&{\tt 2..5873....67}&{\tt 1.355534(011)}&{\tt 0.728205(008)}&{\tt F/1H}\\
{\tt MACHO*04:54:03.4$-$68:52:02}&{\tt 18..2239....43}&{\tt 1.364201(013)}&{\tt 0.801439(021)}&{\tt 1H/2H}\\
{\tt MACHO*05:41:19.1$-$70:16:35}&{\tt 11..9841....64}&{\tt 1.471403(010)}&{\tt 0.729375(008)}&{\tt F/1H}\\
{\tt MACHO*05:27:59.9$-$69:37:20}&{\tt 77..7673...103}&{\tt 1.546895(023)}&{\tt 0.724540(012)}&{\tt F/1H}\\
{\tt MACHO*05:15:31.3$-$68:50:11}&{\tt 79..5748...100}&{\tt 1.569342(023)}&{\tt 0.727638(011)}&{\tt F/1H}\\
&&&&\\
{\tt MACHO*05:36:54.7$-$70:08:10}&{\tt 81..9117...120}&{\tt 1.785987(023)}&{\tt 0.727440(013)}&{\tt F/1H}\\
{\tt MACHO*05:49:14.2$-$71:09:53}&{\tt 15.11158....33}&{\tt 1.829270(018)}&{\tt 0.724838(010)}&{\tt F/1H}\\
{\tt MACHO*05:33:56.7$-$68:53:29}&{\tt 82..8652....67}&{\tt 1.884106(038)}&{\tt 0.722736(017)}&{\tt F/1H}\\
{\tt MACHO*04:49:10.1$-$67:51:53}&{\tt 47..1407....18}&{\tt 1.894536(040)}&{\tt 0.723646(017)}&{\tt F/1H}\\
{\tt MACHO*04:51:31.3$-$67:48:45}&{\tt 47..1771....30}&{\tt 1.896089(033)}&{\tt 0.723350(014)}&{\tt F/1H}\\
{\tt MACHO*05:00:55.1$-$69:16:32}&{\tt 18..3322...549}&{\tt 1.904173(019)}&{\tt 0.723137(008)}&{\tt F/1H}\\
{\tt MACHO*04:54:55.0$-$69:14:12}&{\tt 18..2354....83}&{\tt 2.214457(038)}&{\tt 0.721260(016)}&{\tt F/1H}\\
{\tt MACHO*05:36:31.4$-$69:28:16}&{\tt 81..9127....78}&{\tt 2.726629(116)}&{\tt 0.716381(036)}&{\tt F/1H}\\
{\tt MACHO*05:06:02.7$-$68:06:02}&{\tt 19..4186...876}&{\tt 2.782543(027)}&{\tt 0.719885(011)}&{\tt F/1H}\\
{\tt MACHO*05:23:13.8$-$70:13:46}&{\tt 6..6937....46}&{\tt 2.798537(130)}&{\tt 0.718560(035)}&{\tt F/1H}\\
&&&&\\
{\tt MACHO*05:27:06.1$-$70:19:38}&{\tt 7..7541....17}&{\tt 2.849367(026)}&{\tt 0.713247(010)}&{\tt F/1H}\\
{\tt MACHO*05:26:18.8$-$70:31:56}&{\tt 7..7417....22}&{\tt 2.935517(022)}&{\tt 0.717742(007)}&{\tt F/1H}\\
{\tt MACHO*05:33:39.4$-$69:54:55}&{\tt 81..8636....55}&{\tt 2.937037(030)}&{\tt 0.715276(011)}&{\tt F/1H}\\
{\tt MACHO*05:15:31.1$-$69:18:04}&{\tt 79..5741....67}&{\tt 3.170829(048)}&{\tt 0.716971(011)}&{\tt F/1H}\\
{\tt MACHO*05:21:54.4$-$69:23:04}&{\tt 78..6708....26}&{\tt 3.321481(020)}&{\tt 0.706737(006)}&{\tt F/1H}\\
{\tt MACHO*05:29:36.0$-$69:40:28}&{\tt 77..8035....22}&{\tt 3.405098(050)}&{\tt 0.715876(012)}&{\tt F/1H}\\
{\tt MACHO*05:27:15.8$-$69:43:43}&{\tt 77..7550....42}&{\tt 3.433695(026)}&{\tt 0.710865(008)}&{\tt F/1H}\\
{\tt MACHO*05:31:09.0$-$70:05:11}&{\tt 77..8271....26}&{\tt 3.446769(025)}&{\tt 0.712526(007)}&{\tt F/1H}\\
{\tt MACHO*05:35:56.8$-$70:04:51}&{\tt 81..8997....28}&{\tt 3.455026(042)}&{\tt 0.714131(011)}&{\tt F/1H}\\
{\tt MACHO*05:34:59.5$-$71:12:35}&{\tt 14..8859....14}&{\tt 3.641905(038)}&{\tt 0.717143(010)}&{\tt F/1H}\\
&&&&\\
{\tt MACHO*05:30:59.9$-$69:49:17}&{\tt 77..8154....22}&{\tt 3.685674(030)}&{\tt 0.716438(007)}&{\tt F/1H}\\
{\tt MACHO*05:06:29.3$-$68:54:20}&{\tt 1..4174....23}&{\tt 3.834480(042)}&{\tt 0.712401(010)}&{\tt F/1H}\\
{\tt MACHO*05:03:58.5$-$69:25:38}&{\tt 1..3804....26}&{\tt 3.892839(043)}&{\tt 0.718364(010)}&{\tt F/1H}\\
{\tt MACHO*05:20:07.3$-$70:04:09}&{\tt 78..6456....11}&{\tt 4.840916(055)}&{\tt 0.713243(012)}&{\tt F/1H}\\
{\tt MACHO*05:20:07.1$-$70:04:08}&{\tt 6..6456..4343}&{\tt 4.840942(169)}&{\tt 0.713247(034)}&{\tt F/1H}\\
&&&&\\
\hline
\end{tabular}
\end{center}

\section{Work in Progress}

\vfill

\begin{thebibliography}{99}{\baselineskip 0.4cm
\bibitem{AlcockBC} Alcock, C., {\it et al.} 1995. \aj\, {\bf 109}, 1653
\bibitem{AlcockRRd1} Alcock, C., {\it et al.} 1996. \aj\, {\bf 111}, 1146
\bibitem{AlcockRCrB} Alcock, C., {\it et al.} 1996. \aj\, (in press)
\bibitem{AlcockRRd2} Alcock, C., {\it et al.} 1997. \apj\, (submitted)
\bibitem{AlcockBC2} Alcock, C., {\it et al.} 1997. \aj\, (in preparation)
\bibitem{Beaulieu1995} Beaulieu, J.P. {\it et al.} 1995. \aa, 303, 137
\bibitem{SL1981} Simon, N.R., \& Lee, A.S. 1981. \apj, 248, 291
\bibitem{Welch1995} Welch, D.L., {\it et al.} 1995. in {\it Astrophysical
Applications of Stellar Pulsation}, {\it ASP Conf.\ Series}, 83, 282
}
\end{thebibliography}
\end{document}